\documentclass{ws-procs9x6}

\begin{document}

\title{Extraction of \boldmath$a_{nn}$ from \boldmath$\pi^-d\to nn\gamma$}

\author{A. G{\aa}rdestig}

\address{Department of Physics and Astronomy,
University of South Carolina,\\
Columbia, SC 29210, U.S.A.\\
E-mail: anders@physics.sc.edu}

\begin{abstract}
I present a calculation of the $\pi^-d\to nn\gamma$ reaction to third order in
chiral perturbation theory.
The short-distance physics of this reaction can be constrained by relating it
to several important low-energy weak reactions.
The theoretical error in $a_{nn}$ extracted from this reaction can thus 
be reduced by a factor larger than three to $\pm0.05$~fm.
\end{abstract}


\bodymatter
\mbox{}\\[1ex]
Since there are no neutron targets, the neutron-neutron scattering length 
($a_{nn}$) can only be accessed using indirect methods, with neutrons detected 
in phase space regions sensitive to $a_{nn}$.
Measurements of $a_{nn}$ using $nd\to nnp$ suffer from unresolved
discrepancies between experiments~\cite{Calvin}.
Experiments using the final state of $\pi^-d\to nn\gamma$ are more consistent 
and dominate the presently accepted value of 
$a_{nn}=-18.59\pm0.4$~fm~\cite{LAMPF,NNreview}.
The latter extractions used theory models by Gibbs, Gibson, and 
Stephenson and de~T\'eramond et al.~\cite{GGSdeTeramond}, both with a 
theoretical error of $\pm0.3$~fm.
I will show how chiral perturbation theory can reduce this error 
considerably~\cite{GP1,GP2,AG1}.

The $\pi^-d\to nn\gamma$ reaction is dominated by well-known single-nucleon 
photon-pion amplitudes.
Two-nucleon diagrams occur first at $\mathcal{O}(Q^3)$ (where $Q\sim m_\pi$ is 
a small energy/momentum) but nevertheless cause a small, but significant, 
change in the shape of the $\pi^-d\to nn\gamma$ spectrum.
The wave functions for the bound and scattering states are calculated from 
the known asymptotic behavior, integrated in from $r=\infty$ using the
one-pion-exchange potential.
At short distances, the unknown short-distance physics is parametrized and
regularized by matching the wave functions at some radius $r=R$ 
($1.4$~fm~$<R<3$~fm) to a spherical well solution for $r<R$.
Details about the calculation and the wave functions can be
found in Ref.~\refcite{GP1}.

\begin{figure}[Ht]
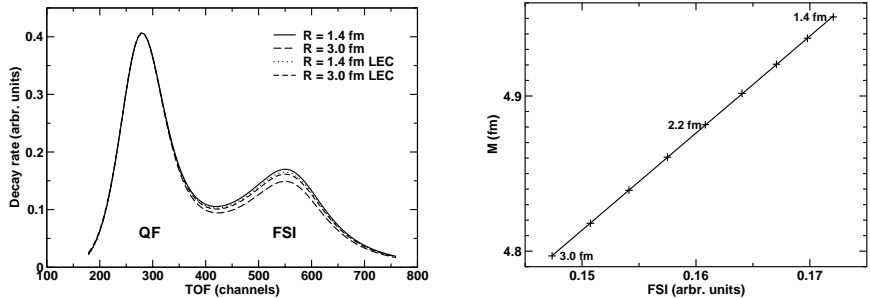

\psfig{file=gardestig_1_a.eps,width=2.2in,clip}%
\hfill\psfig{file=gardestig_1_b.eps,width=2in,clip}
\caption{(Left panel) Neutron time-of-flight spectrum for $\pi^-d\to nn\gamma$,
without and with the LEC as indicated.
The labels indicate where the reaction is dominated by quasi-free (QF) and
final-state-interaction (FSI) kinematics.
(Right panel) The relation between the GT matrix element and the FSI peak 
height, for varying $R$ as indicated.}
\label{fig:TOFGP}
\end{figure}

The result at $\mathcal{O}(Q^3)$ is sensitive to the value of $R$, indicating 
that some unknown short-distance physics is at play 
(left panel of Fig.~\ref{fig:TOFGP}).
As shown in Refs.~\refcite{GP2,AG1}, the $\pi^-d\to nn\gamma$ reaction is 
sensitive to the same short-distance physics as several important weak 
reactions, e.g., solar $pp$ fusion and tritium beta decay.
Calculations confirm this.
Figure~\ref{fig:TOFGP} (right panel) shows a linear relation between the 
$\pi^-d\to nn\gamma$ FSI peak height and the Gamow-Teller (GT)
matrix element of $pp$ fusion. 
This indicates that both reactions can be
simultaneously renormalized by \emph{one} low-energy constant (LEC). 
The same LEC appears in the chiral three-nucleon force (3NF) with similar 
kinematics, enabling extracting (part of) the 3NF from two-nucleon systems.

Once the LEC is included the theoretical error due to short-distance physics
can be reduced significantly, as is obvious from Fig.~\ref{fig:TOFGP}.
Including other sources of errors~\cite{GP1}, the theoretical
error in $a_{nn}$ from fitting the entire spectrum is $\pm0.3$~fm, while 
fitting only the FSI peak gives $\pm0.05$~fm~\cite{GP2}.
For further details and discussions about future work, 
see Refs.~\refcite{GP1,GP2,AG1}.

This work was supported in part by DOE grant No.\ DE-FG02-93ER40756 and 
NSF grant No.\ PHY-0457014.

\end{document}